\documentclass{aa}  

\usepackage{graphicx}
\usepackage{subcaption}
\usepackage{array}
\usepackage{txfonts}
\usepackage{hyperref}
\begin{document}

   \title{J-VAR: Analysis of RR Lyrae light curves in seven optical bands}
   \titlerunning{J-VAR: RR Lyrae in seven optical bands}
   \author{
S.~Kulkarni\inst{1}\thanks{E-mail: skulkarni@cefca.es}
\and H.~V\'azquez~Rami\'o\inst{1,2}
\and C.~L\'opez-Sanjuan\inst{1,2}
\and S.~Pyrzas\inst{1}
\and A.~Ederoclite\inst{1,2}
\and F.~Jim\'enez-Esteban\inst{10}
\and A.~J.~Cenarro\inst{1,2}
\and A.~Mar\'{\i}n-Franch\inst{1,2}
\and J.~Alcaniz\inst{5}
\and R.~E.~Angulo\inst{8,9}
\and D.~Crist\'obal-Hornillos\inst{1}
\and R.~A.~Dupke\inst{5,6}
\and C.~Hern\'andez-Monteagudo\inst{3,4}
\and M.~Moles\inst{1}
\and L.~Sodr\'e~Jr.\inst{7}
\and J.~Varela\inst{1}
}
\authorrunning{Kulkarni et. al.}
\institute{
Centro de Estudios de F\'{\i}sica del Cosmos de Arag\'on (CEFCA),
Plaza San Juan 1, 44001 Teruel, Spain
\and
Unidad Asociada CEFCA-IAA, CEFCA, Unidad Asociada al CSIC por el IAA y el IFCA,
Plaza San Juan 1, 44001 Teruel, Spain
\and
Instituto de Astrof\'{\i}sica de Canarias,
La Laguna, 38205 Tenerife, Spain
\and
Departamento de Astrof\'{\i}sica, Universidad de La Laguna,
38206 Tenerife, Spain
\and
Observat\'orio Nacional - MCTI (ON),
Rua Gal. Jos\'e Cristino 77, S\~ao Crist\'ov\~ao,
20921-400 Rio de Janeiro, Brazil
\and
University of Michigan, Department of Astronomy,
1085 South University Ave., Ann Arbor, MI 48109, USA
\and
Instituto de Astronomia, Geof\'{\i}sica e Ci\^encias Atmosf\'ericas,
Universidade de S\~ao Paulo,
05508-090 S\~ao Paulo, Brazil
\and
Donostia International Physics Centre (DIPC),
Paseo Manuel de Lardizabal 4,
20018 Donostia-San Sebasti\'an, Spain
\and
IKERBASQUE, Basque Foundation for Science,
48013 Bilbao, Spain
\and
Centro de Astrobiolog\'{\i}a (CAB), CSIC-INTA,
Camino Bajo del Castillo s/n,
E-28692 Villanueva de la Ca\~nada, Madrid, Spain
}

   \date{Received \textbf{XXX}; accepted \textbf{XXX}}

% \abstract{}{}{}{}{} 
% 5 {} token are mandatory
 
  \abstract
   {
   RR Lyrae stars, with their accurate period and amplitude measurements, constrain stellar evolution and map Galactic structure. The Javalambre VARiability (J-VAR) survey is the time-domain extension of the Javalambre Photometric Local Universe survey, which provides time-series data across seven optical bands, including $gri$ and four medium and narrow bands.
   }
   {
   Our goal is to construct and analyze light curves for RR Lyrae stars identified in the J-VAR's first data release using the \textit{Gaia} third data release (DR3) Variable Stars catalog as a reference.
   }
   { 
   The light curves of $315$ RR Lyrae were analyzed by fitting templates from the Sloan Digital Sky Survey Multiband Template Library. The periods and amplitudes for the seven bands in J-VAR were independently obtained from the best-fitted templates.
   }
   {
    The J-VAR periods show strong agreement with \textit{Gaia} DR3 values. The Bailey diagram for each J-VAR filter shows larger pulsation amplitudes at bluer wavelengths. Amplitudes, after normalizing by the $r$-band amplitude, show an exponential trend, with the bluer J-VAR filter centered at $395~\text{nm}$ having twice the amplitude of the reddest J-VAR passband at $861~\text{nm}$. The normalized amplitudes of the RR Lyrae stars from {\it Gaia} and the Zwicky Transient Facility are consistent with the J-VAR trend. Finally, the SDSS templates derived from broadbands also provide a proper description for the medium and narrow band light curves.
   }
   {
   The J-VAR RR Lyrae catalog offers reliable pulsation parameters and light curves in seven optical filters, allowing the systematic study of amplitude trends from $395~\text{nm}$ to $860~\text{nm}$.
   }

   \keywords{Stars: variables: RR Lyrae - Surveys: J-VAR - Techniques: photometric - methods: statistical}
\maketitle
  
%
%-------------------------------------------------------------------
\section{Introduction}
\label{sec:Introduction}
Variable stars are among the most valuable tools in astronomy. These stars serve as reliable standard candles for determining cosmic distances \citep{CacciariClementini2003} and offer critical insights into stellar evolution and Galactic structure \citep{Catelan2009, catelan2015}. As excellent tracers of old stellar populations, variable stars such as RR Lyrae, located in the halo of the Milky Way, are ideal for studying their formation and evolution \citep{Pietrukowicz2016, Prudil2022}. Furthermore, RR Lyrae stars having pulsation periods in fundamental frequency are classified as RRab subtype; those having periods in the first overtones are classified as RRc subtype, and those having periods in both frequencies are classified as RRd subtype stars. 

Recent large-scale surveys have revolutionized our understanding of RR Lyrae stars. The \textit{Gaia} (\textit{Gaia} Collaboration \citeyear{GaiaMission16}), particularly the third data release (DR3; \textit{Gaia} Collaboration \citeyear{GaiaDR3_2022}), provide precise astrometric, photometric, and spectroscopic data for millions of stars, including RR Lyrae. These massive data have since served as a foundation for several follow-up studies, including Dark Energy Spectroscopic Instrument (DESI) \citep{Medina2025_DESI_Y1_RRL}, among others. In parallel, time-domain surveys such as the Zwicky Transient Facility (ZTF; \citealt{Bellm2019}) have provided optical photometry over large sky areas, enabling the detection and detailed light curve analysis of variable stars, including RR Lyrae, on short timescales.

RR Lyrae stars also exhibit a well-known decrease in pulsation amplitude with increasing wavelength, a trend arising from the reduced contrast between maximum and minimum light at longer wavelengths \citep{Catelan2009}. Most of the photometric studies characterizing this effect have been conducted using broadband filters, which, while efficient for large-scale surveys, blend information from different spectral features. Notable exceptions include narrow-band studies in the Str\"omgren and {\it Caby} photometric systems \citep{baird96,pena09,pena12}, which have demonstrated the diagnostic potential of narrower passbands in probing temperature-sensitive and line-specific variations across the pulsation cycle.

In this context, the Javalambre VARiability Survey first data release (J-VAR DR1; \citealt{Ederoclite_2025}) offers complementary multiband and multiepoch observations in the visible range from $390~\text{nm}$ to $880~\text{nm}$ The J-VAR photometric system (which is a subset of J-PLUS) comprises the \textit{gri} broadbands and four bands located in key stellar features (Tab.~\ref{tab:J-VAR DR1 Bands}): $J0395$ of $10~\text{nm}$ width located at the \ion{Ca}{ii} H+K lines, $J0515$ of $20~\text{nm}$ located at the Mg$b$ triplet, $J0660$ of $14~\text{nm}$ located at H$\alpha$, and $J0861$ of $40~\text{nm}$ located at the calcium triplet. J-VAR DR1 provides photometry in these seven optical bands over 33 epochs, covering $202$ deg$^2$ in the northern sky, with typical single-epoch depths of $m = 19$ mag. This dataset presents a unique opportunity to advance the study of RR Lyrae physics further.

This work analyses the RR Lyrae stars identified in J-VAR DR1 confirmed through cross-matching with the \textit{Gaia} DR3 Variable Star catalog. This ensures a robust classification and provides the periods from the \textit{Gaia} DR3 along with the epochs of maximum brightness. The magnitudes from J-VAR DR1 were fitted using the templates from the Sloan Digital Sky Survey (SDSS; \citealt{york2000}) multiband template library \citep{Sesar2010}. The J-VAR periods and amplitudes are used to study the variation of the RR Lyrae light curve properties along the optical range and to our best knowledge for the first time in the $J0515$, $J0660$ and $J0861$ passbands; being the $J0395$ similar to the $Ca$ filter studied by \citet{baird96}.

This paper is organized as follows. The J-VAR DR1 data, \textit{Gaia} Variable Star catalog, SDSS Multiband Template Library, and ZTF data are presented in Sect.~\ref{sec:Data}, the methodology is described in Sect.~\ref{sec:Methodology}, and the results are detailed in Sect.~\ref{sec:Results}. Furthermore, the discussions and conclusions are in Sects.\ref{sec:Discussion} and \ref{sec:Conclusions}, respectively. Finally, Appendix \ref{App:J-VAR DR1 RR Lyrae Catalog} describes the catalog for light curve analysis of the RR Lyrae stars in J-VAR DR1. The magnitudes presented throughout this work are in the AB scale \citep{oke83}.

% Section 2
\section{Data}\label{sec:Data}

% Figure Bailey Diagram
\begin{figure}
    \centering
    \resizebox{\hsize}{!}{\includegraphics[width=0.5\textwidth]{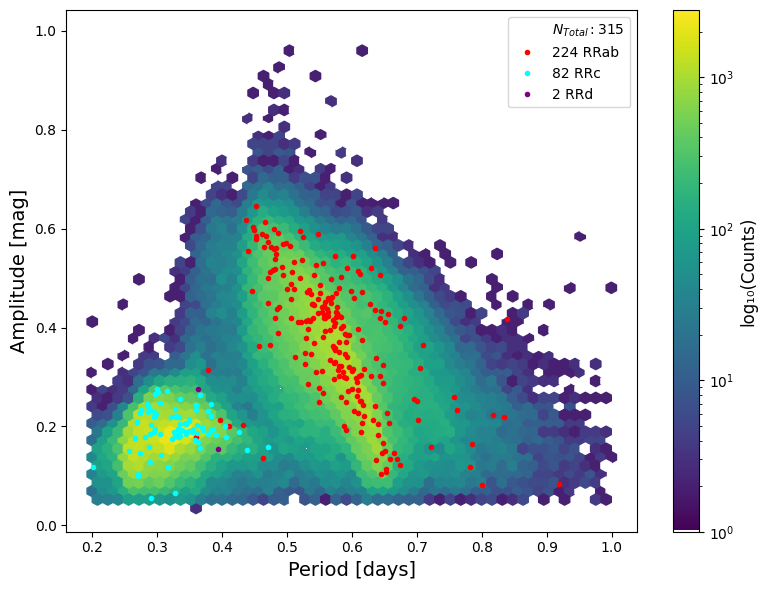}}
    \caption{Bailey Diagram: Density plot for period vs. amplitude distribution of RR Lyrae stars in \textit{Gaia} DR3 Variable Star catalog using log-scaled color bins. The overplotted dots represent the $315$ J-VAR DR1 objects, with RRab (red), RRc (cyan), and RRd (purple) subtypes color-coded.}
    \label{fig:Bailey}
\end{figure}

% Table 1
\begin{table}%[]
    \caption{J-PLUS Photometric Filter Set Adopted by J-VAR}
    \label{tab:J-VAR DR1 Bands}
    \centering
    \begin{tabular}{p{1.5cm}p{1.8cm}p{1.5cm}p{1.5cm}}
    \hline\hline
      Passband & Central Wavelength [\AA] & FWHM [\AA] & Type \\
      \hline 
      $J0395$ & 3950 & 100 & Narrow\\
      $g$ & 4803 & 1409 & Broad\\
      $J0515$ & 5150 & 200 & Medium \\
      $r$ & 6254 & 1388 & Broad\\
      $J0660$ & 6600 & 138 & Narrow\\
      $i$ & 7668 & 1535 & Broad \\
      $J0861$ & 8610 & 400 & Medium\\
      \hline
    \end{tabular}
\end{table}

%Text 
% 2.1
\subsection{J-VAR DR1 Data}
\label{subsec:J-VAR DR1 Data}
J-VAR is conducted from the Observatorio Astrofísico de Javalambre (OAJ, Teruel, Spain; \citealt{oaj}) with the 83 cm Javalambre Auxiliary Survey Telescope (JAST80) and T80Cam, a wide-field camera of 9.2k × 9.2k pixels covering a $2$ deg$^2$ field of view with a pixel scale of $0.55$\arcsec pix$^{-1}$ \citep{t80cam}. J-VAR delivers time-series photometry in seven optical bands across 33 epochs in fields already observed by the single-epoch Javalambre Photometric Local Universe Survey (J-PLUS; \citealt{Cenarro2019}). The observations were taken on non-photometric nights, and the inclusion of four narrow bands enables detailed light curve modeling and spectral energy distribution (SED; \citealt{Pickles1998}, \citealt{Nemec2013}, \citealt{Muraveva2018}) that cover spectral features not typically available in other large-scale variability surveys. 

J-VAR DR1\footnote{\url{https://archive.cefca.es/catalogues/jvar-dr1}} includes light curves for 1,335,279 objects classified as stars or quasars by \texttt{BANNJOS} (Bayesian Artificial Neural Networks for the Javalambre Observatory Surveys; \citealt{bannjos}) in J-PLUS, as detailed in Pyrzas et al., in prep. The data provide a signal-to-noise ratio greater than $10$ for objects brighter than $19$ mag in the $r$-band.

Starting from the J-VAR DR1 \texttt{jvar.Light\_curves} table \citep{Ederoclite_2025} available at CEFCA's catalog portal\footnote{\url{https://archive.cefca.es/catalogues/jvar-dr1}}, some quality cuts were applied. Unreliable photometry estimation is filtered out by excluding magnitudes exceeding $25$ mag in any filter. Measurements with relative errors greater than $20$\% of the magnitude or with absolute errors larger than $0.5$ mag were also discarded. Additionally, custom root-mean-square thresholds (Pyrzas et al., in prep.) were applied to each band to remove outliers.

\subsection{\textit{Gaia} DR3 Variable Star Catalog}
\label{subsec:Gaia DR3 Variable Star Catalog} % 2.2
\textit{Gaia} is a space observatory dedicated to mapping over 1.8 billion stars in the Milky Way. It monitors stellar positions, motions, and brightness variations, enabling large-scale studies of variable stars \citep{GaiaCollaboration2022}.

The \textit{Gaia} DR3 Variable Star catalog $I/358/vrrlyr$ \citep{vrrlyr_catalog_2023} has $271,779$ variable stars identified as RR Lyrae. It comprises $175,350$ RRab, $94,422$ RRc, and $2,007$ RRd subtypes. The catalog provides information for photometry in \textit{Gaia} $G$-band, periods in fundamental frequency and first overtones, and epochs of maximum light. 

\subsection{Data Cross-matching}% 2.3
\label{subsec:Data Cross-matching and Parameter Estimation}
A Python-based pipeline was developed to perform a cross-match between objects in the J-VAR DR1 data and RR Lyrae stars identified in the \textit{Gaia} DR3 Variable Star catalog. While J-VAR provides object lists and multiband photometry, the \textit{Gaia} DR3 Variable Star catalog supplies variability information, including periods (fundamental and first overtone) and epochs of maximum brightness. The cross-match was carried out using the \texttt{gaia\_sid} identifiers available in both datasets, rather than a purely positional match. This procedure resulted in a combined J-VAR DR1 + \textit{Gaia} DR3 sample of 315 RR Lyrae stars. Of these, 308 could be assigned to specific subtypes ($224$ RRab, $82$ RRc, and $2$ RRd), while the remaining $7$ stars lacked well-defined periods in \textit{Gaia} DR3 and were therefore left unclassified.

To construct light curves, the Modified Julian Dates (MJD) in J-VAR DR1 were converted to Barycentric Julian Dates (BJD) using functions from \texttt{Astropy} \citep{AstropyCollaboration2013, astropy:2018, astropy:2022}, which account for the Earth’s motion relative to the barycenter of the Solar System. The transformation incorporated the coordinates of the observation site, latitude $40.0451^\circ$, longitude $-1.0005^\circ$, and elevation $1957~\text{m}$ using the \texttt{EarthLocation} module in \texttt{Astropy} to ensure precise timing corrections.

Fig.~\ref{fig:Bailey} shows the Bailey Diagram \citep{Bailey1981}, which plots \textit{Gaia} periods versus the \textit{Gaia} amplitudes, for the entire {\it Gaia} DR3 catalog and the $315$ RR Lyrae in the J-VAR DR1 catalog. The amplitudes were assumed as halves of the peak-to-peak magnitudes in the \textit{Gaia} $G$-band. The alignment of objects in J-VAR DR1 with the sources in the \textit{Gaia} DR3 Variable Star catalog is confirmed, providing a representative subset of the initial \textit{Gaia} sample. The gap between the RRab and RRc populations may be partly influenced by residual systematics in the \textit{Gaia}$-$J-VAR amplitude calibration \citep{Clementini_2023}, or it may simply reflect the intrinsic Oosterhoff dichotomy (\citealt{Oosterhoff_1939}, \citealt{catelan2015}), which naturally produces distinct RRab and RRc loci.

\subsection{SDSS Multiband Template Library}%2.4
\label{subsec:SDSS Multiband Template Library}
The SDSS is a wide-field imaging and spectroscopic survey that has significantly advanced our understanding of the Milky Way and the distant universe. The SDSS Multiband Template Library for RR Lyrae stars, as presented in \cite{Sesar2010}, provides a set of well-defined empirical light curve templates based on observations of $483$ RR Lyrae stars ($379$ RRab and $104$ RRc stars) from SDSS Stripe 82. These templates were constructed using a large sample of well-sampled multiband light curves in the $ugriz$ photometric system, with a median of $30$ observations per band for objects with $g<21$ mag. The template families found in that work show a variety of shapes of the light curves. The study demonstrated that RR Lyrae stars, particularly RRab and RRc subtypes, exhibit characteristic light curve shapes that can be effectively modeled using these templates. The library improves period determination, classification accuracy, and distance estimates for RR Lyrae stars, aiding studies of the Galactic structure, especially in the halo. The SDSS template library was used to fit the J-VAR DR1 multiband light curves of each RR Lyrae star.

\subsection{The ZTF Catalog}\label{subsec:The ZTF Catalog}
To complement the J-VAR and Gaia photometry, the light-curve information from the Zwicky Transient Facility (ZTF; \citealt{Masci2019}) was also used. ZTF provides multi-epoch photometry in the $g$ and $r$ bands, with passband profiles broadly similar to the corresponding J-VAR filters. Amplitude measurements for RR Lyrae stars were taken from the ZTF DR3 catalog (J/ApJS/249/18)\footnote{\url{https://vizier.cds.unistra.fr/viz-bin/VizieR-3?-source=J/ApJS/272/31}} 
\citep{Chen2020_ZTF_Catalog}. These values are used for cross-survey comparisons in Sect.~\ref{subsec:Amplitude Trends}.

%%%
% 3
\section{Methodology}\label{sec:Methodology}

% Figure Light curves
\begin{figure*}%[ht]
  \centering
  \begin{subfigure}{0.33\textwidth}
    \centering
    \includegraphics[width=\linewidth]{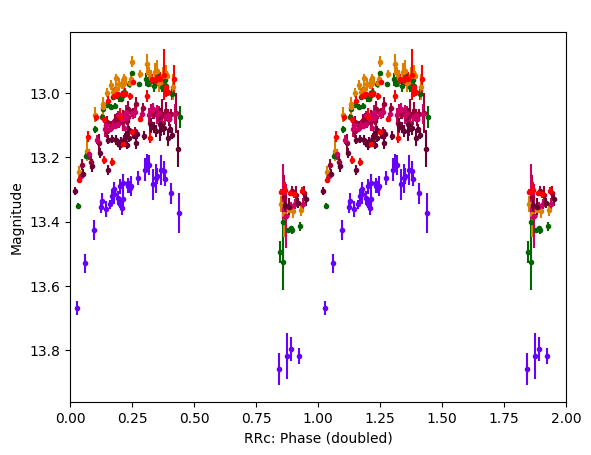}
    \caption{J-VAR DR1 OBJECT ID: 90815-43021}
    \label{fig:J-VAR DR1 OBJECT ID: 90815-43021}
  \end{subfigure}
  \begin{subfigure}{0.33\textwidth}
    \centering
    \includegraphics[width=\linewidth]{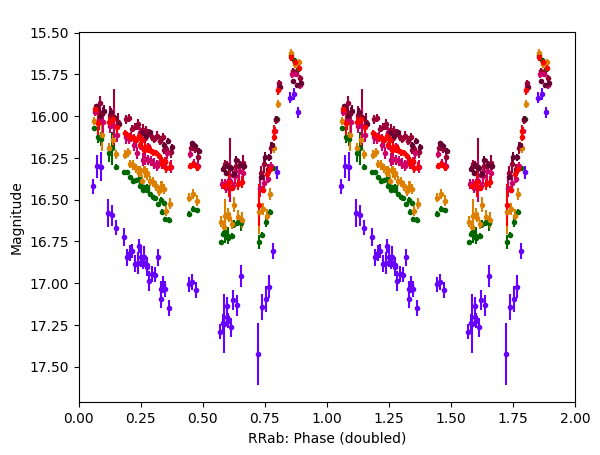}
    \caption{J-VAR DR1 OBJECT ID: 87675-98430}
    \label{fig:J-VAR DR1 OBJECT ID: 87675-98430}
  \end{subfigure}
  \begin{subfigure}{0.33\textwidth}
    \centering
    \includegraphics[width=\linewidth]{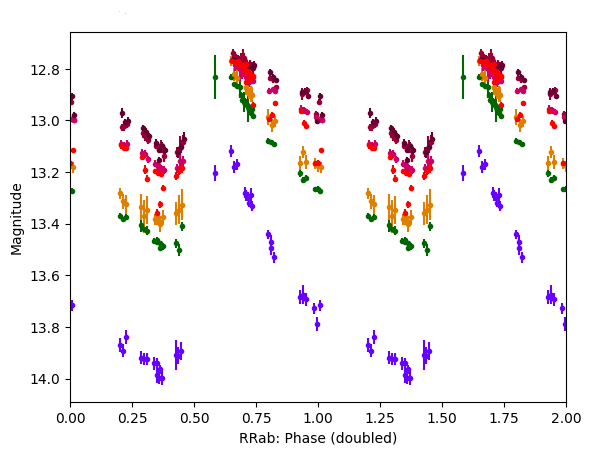}
    \caption{J-VAR DR1 OBJECT ID: 90007-23972}
    \label{fig:J-VAR DR1 OBJECT ID: 90007-23972}
  \end{subfigure}
  \vspace{0.5cm} % Add vertical space between rows
  \begin{subfigure}{0.33\textwidth}
    \centering
    \includegraphics[width=\linewidth]{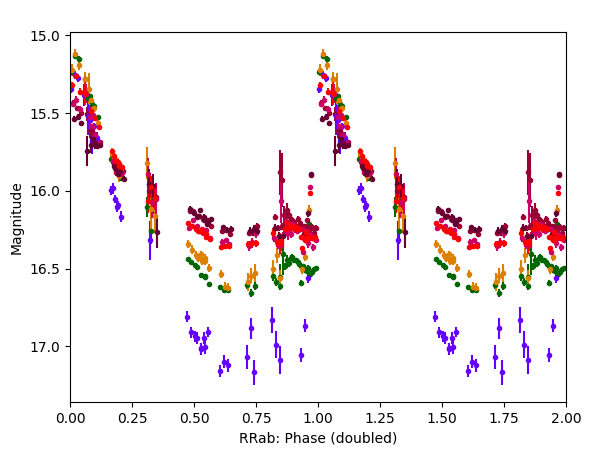}
    \caption{J-VAR DR1 OBJECT ID: 85953-3293}
    \label{J-VAR DR1 OBJECT ID: 85953-3293}
  \end{subfigure}
  \begin{subfigure}{0.33\textwidth}
    \centering
    \includegraphics[width=\linewidth]{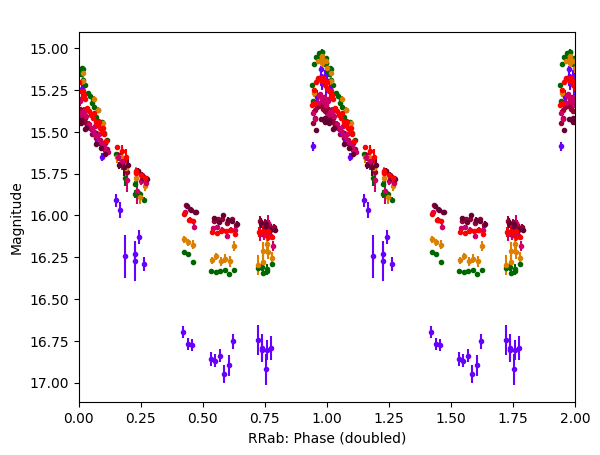}
    \caption{J-VAR DR1 OBJECT ID: 86612-10980}
    \label{fig:J-VAR DR1 OBJECT ID: 86612-10980}
  \end{subfigure}
  \begin{subfigure}{0.33\textwidth}
        \centering
        \includegraphics[width=\linewidth]{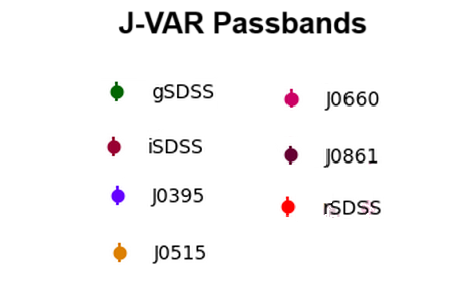}
        %\caption{Overall heading panel}
        \label{fig:bands}
  \end{subfigure}
  \caption{Light curves: The figure shows phase-folded light curves for five example RR Lyrae in the J-VAR DR1 + \textit{Gaia} DR3 catalog. A RR Lyrae star of RRc subtype (\textit{Upper Left panel}) and four RR Lyrae stars of RRab subtype (\textit{Remaining panels}) are shown. The unique identification numbers for objects in J-VAR DR1 are provided.}
  \label{fig:Light curves}
\end{figure*}

% figure Template fitting
\begin{figure*}
    \centering
    \begin{subfigure}{0.48\textwidth}
        \centering
        \includegraphics[width=\textwidth]{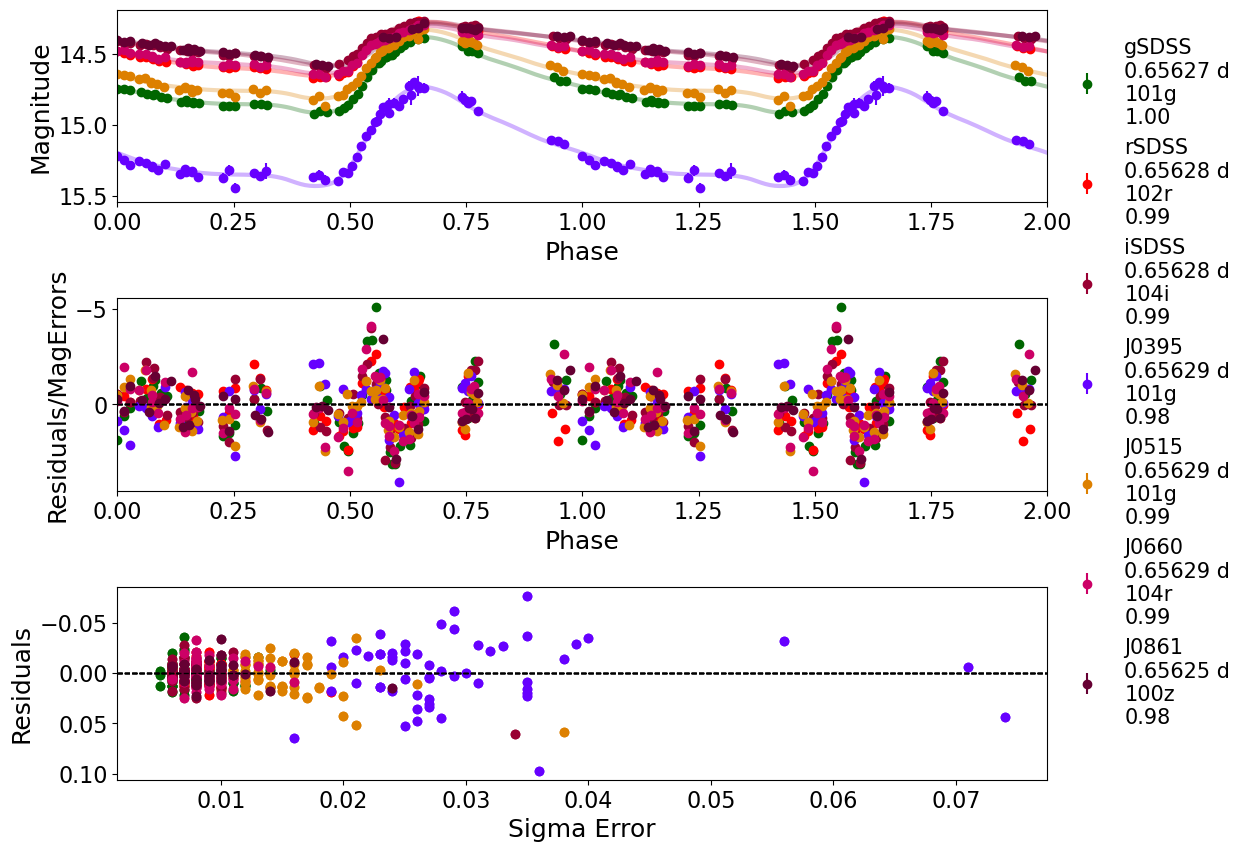}
        \caption{J-VAR DR1 OBJECT ID: 92837-7186, Subtype: RRab}
        \label{fig:J-VAR DR1 OBJECT ID: 92837-7186, Subtype: RRab}
    \end{subfigure}
    \begin{subfigure}{0.48\textwidth}
        \centering
        \includegraphics[width=\textwidth]{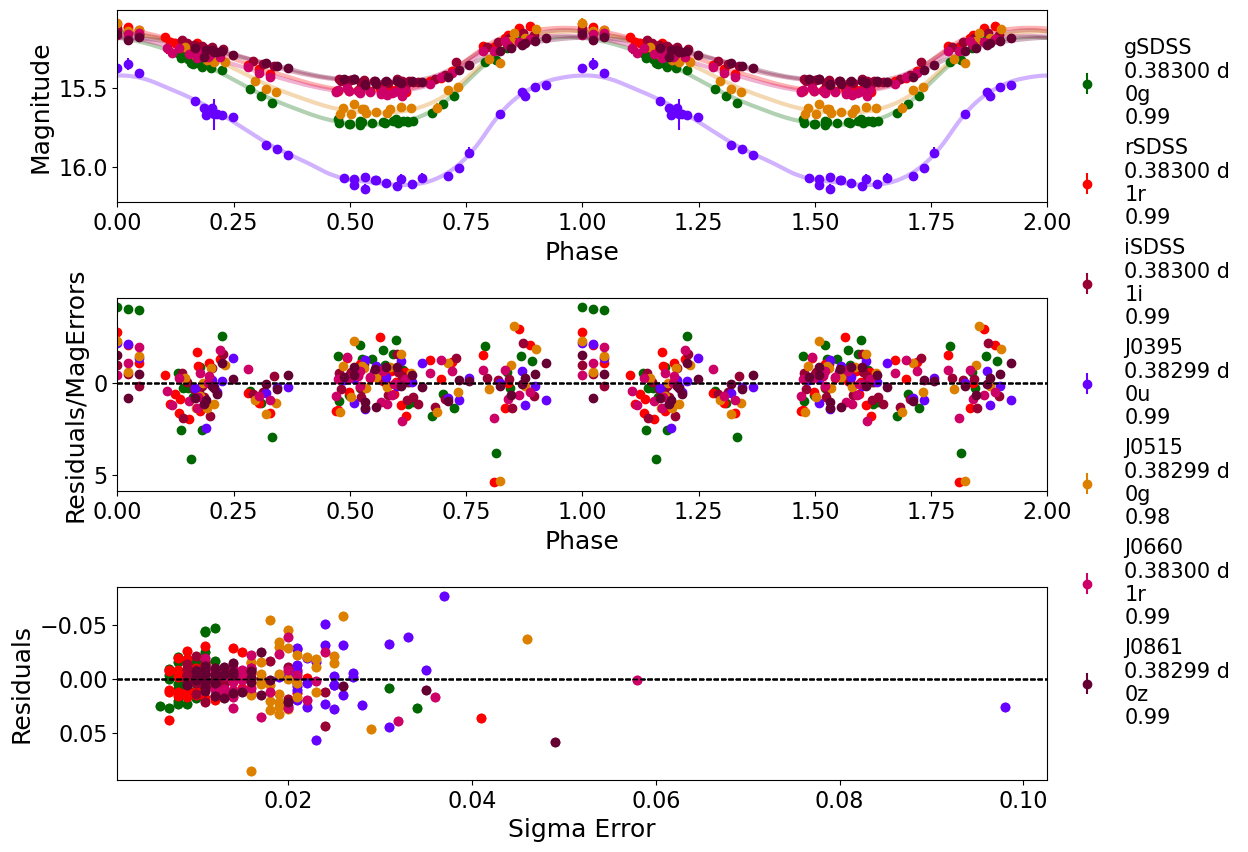}
        \caption{J-VAR DR1 OBJECT ID: 84800-11425, Subtype: RRc}
        \label{fig:J-VAR DR1 OBJECT ID: 84800-11425, Subtype: RRc}
    \end{subfigure}
 \caption{Light curve analysis in J-VAR DR1 for two representative examples. {\it Top panels}: The light curve analysis by fitting the best-fitted templates. {\it Middle panels}: Residuals normalized by magnitude errors plotted against phase. {\it Bottom panels}: The Residuals plotted against sigma errors in magnitudes. A solid black line in the plot marks the zero reference point. The legends in the right of each plot describe the bands (e.g., $g$-band), the periods of the best-fitted templates (e.g., $0.65627$ d, also known as J-VAR periods), the number and band of the best-fitted template (e.g., $101g$), and the $\chi\textsuperscript{2}$ test (e.g., $1.00$). The J-VAR DR1 unique identification numbers are also provided.}
 \label{fig:template_fit_subpanels}
\end{figure*}
    
%%% TEXT %%% 
\subsection{Phase-folded Light Curves}%3.1
\label{subsec:Light Curve Creation}
Following the cross-match between the J-VAR DR1 data and the \textit{Gaia} DR3 Variable Star catalog, a combined catalog was created. However, this initial version did not include phase-folded light curves. To address this, an additional step was implemented using a custom pipeline that phase-folds the light curves from phase $0$ to phase $1$ using a reference period, normalized over the pulsation cycle. Each light curve is plotted twice ($0-2$ in phase) to display the variability pattern. Fig.~\ref{fig:Light curves} shows five representative examples generated using this method and the {\it Gaia} periods for reference.

\subsection{Parameter Estimation}%3.2
\label{subsec:Parameter Estimation}
\begin{table}%[] % Table 2
     \caption{J-VAR Band Configurations with respect to SDSS Template Library Bands}
    \label{tab:configurations}
    \centering
    \begin{tabular}{p{2.5cm}m{2.5cm}}
    \hline\hline
       J-VAR  & SDSS \\
       \hline
       $J0395$ & $ug$ \\
       $g$ & $g$  \\
       $J0515$ & $g$  \\
       $r$ & $r$  \\
       $J0660$& $r$  \\
       $i$ & $i$  \\
       $J0861$ & $z$  \\
       \hline
    \end{tabular}
\end{table}

%%% Text
The light curves of the $315$ RR Lyrae stars in the J-VAR DR1 + \textit{Gaia} DR3 catalog were analyzed using a robust template-fitting pipeline. The workflow integrates widely used tools: \texttt{Astropy} for time and coordinate transformations, \texttt{Gatspy} \citep{Vanderplas2015} for period determination and multiband light curve modeling, \texttt{NumPy} \citep{Harris_2020} for numerical calculations, and \texttt{astroML} \citep{Vanderplas2012} for statistical methods, including period refinement. Several \texttt{Gatspy} modules were modified to accelerate period determination through parallelization and to enable the retrieval of the best-fitting template label.

Tab.~\ref{tab:configurations} outlines the band configurations of J-VAR DR1 and SDSS used for the template fitting. The broad bands $gri$ were aligned with their direct SDSS counterparts, while the medium and narrow bands were mapped to the closest SDSS filters by effective wavelength. The script fits each light curve over a period range from $0-1$ days using the \texttt{RRLyraeTemplateModeler}. Fig.~\ref{fig:template_fit_subpanels} shows the light curve analysis for two representative RR Lyrae stars in the catalog.

The template-fitting process provided independent parameter estimates across all seven optical bands. These include the identifier of the best-fitting template, periods of the best-fitted templates (also known as J-VAR periods), and the corresponding $\chi^{2}$ value, which is reported for completeness but not further used in this study. The uncertainties on the J-VAR periods are extremely small ($\lesssim 10^{-6}$--$10^{-5}$~days), and are therefore negligible for this analysis. As a part of the catalog, J-VAR amplitudes, defined as half of the peak-to-peak magnitudes, were also recorded.
%%%%%%%
%\ Section 4
\section{Results}
\label{sec:Results}
%% Figures
% figure Bailey J-VAR DR1
\begin{figure*}
    \centering
    \begin{subfigure}{0.48\textwidth}
     \centering
        \includegraphics[width=\textwidth]{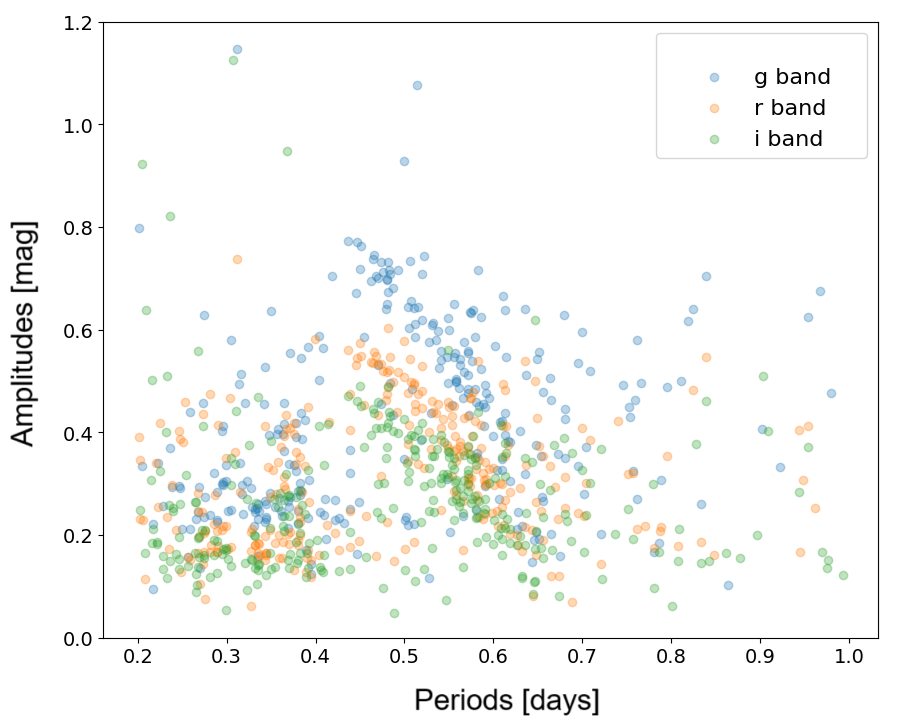}
        \label{}
    \end{subfigure}
    \begin{subfigure}{0.48\textwidth}
        \centering
        \includegraphics[width=\textwidth]{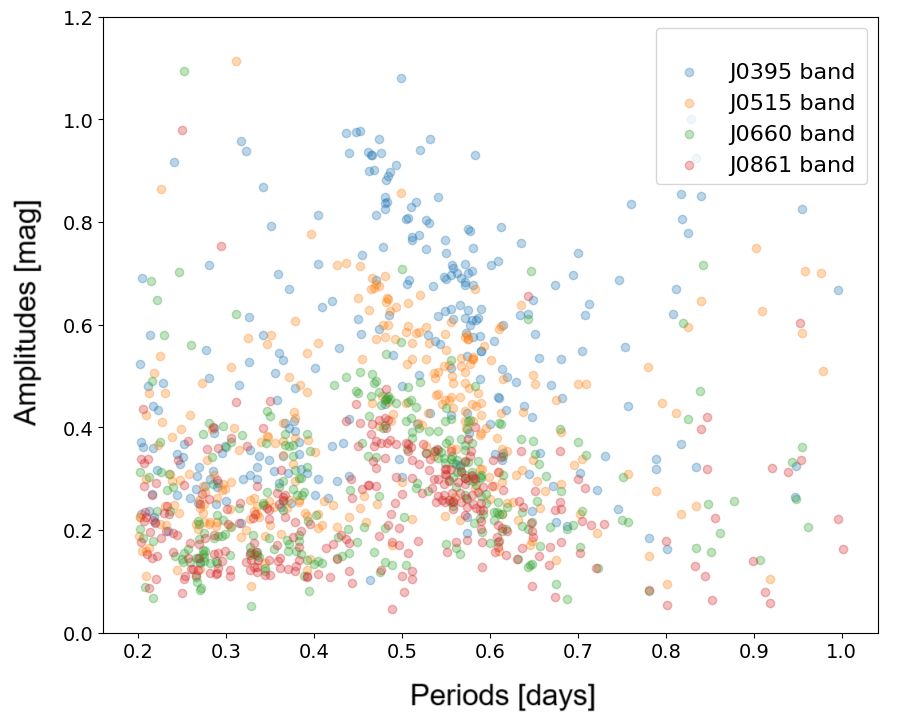}
        \label{fig:}
    \end{subfigure}
 \caption{Bailey Diagram with J-VAR periods and J-VAR amplitudes. {\it Left panel}: $gri$ bands. {\it Right panel}: $J0395$, $J0515$, $J0660$, and $J0861$ bands.}
 \label{fig:Bailey_JVAR}
\end{figure*}

% figure Normalized Amplitude vs Wavelengths
\begin{figure}
    \centering
    \resizebox{\hsize}{!}{\includegraphics[width=0.5\textwidth]{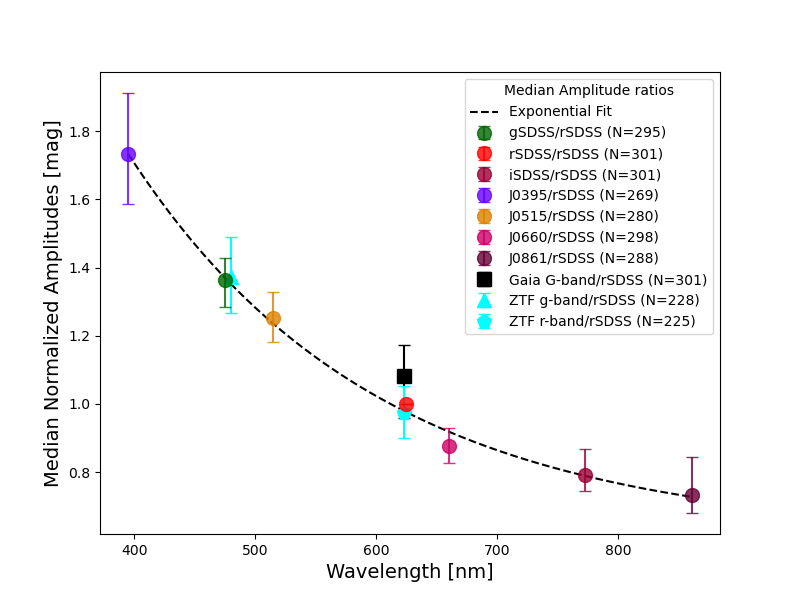}}
    \caption{Normalized amplitudes vs. wavelength. The points represent the medians of the J-VAR amplitudes for each band normalized by that in the $r$-band. The figure also shows the \textit{Gaia} $G$-band and the ZTF $g$ and $r$ bands (normalized by the J-VAR $r$-band) for comparison and were not included in the fit. The dashed black line indicates the exponential fit. Error bars correspond to the 25\textsuperscript{th} ($p_{25}$) and 75\textsuperscript{th} ($p_{75}$) percentiles. N denotes the number of valid amplitude ratios computed per band.}
    \label{fig:normApli_vs_wavelengths}
\end{figure}

% figure Residuals Histogram
\begin{figure}%[t]
    \centering
    \resizebox{\hsize}{!}{\includegraphics[width=0.5\textwidth]{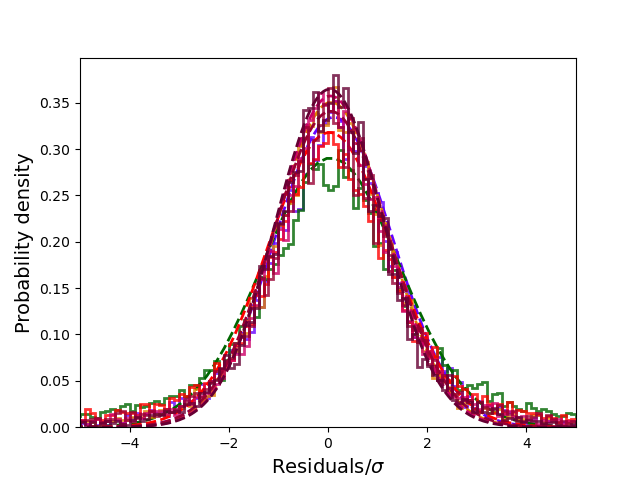}}
    \caption{Histogram of Residuals/$\sigma$: The figure shows a histogram for the ratio of residuals to the errors in the magnitudes with Gaussian fits per band.}
    \label{fig:hist_residuals}
\end{figure}

%% TEXT %%
\subsection{Bailey diagram}
\label{subsec:Bailey Diagram}  % 4.1
The Bailey diagram with J-VAR amplitudes and J-VAR periods (Figure \ref{fig:Bailey_JVAR}) reveals a clear separation between RRab and RRc subtypes across both the broad ($gri$) and narrow bands. Moreover, the characteristic locus for RRab stars is more distinctly visible compared to that for RRc stars. The J-VAR amplitudes in the bluer passbands are larger than in the redder ones, both for the broad and the medium-to-narrow filters, following the known trend with wavelength \citep[e.g.][]{Catelan2009}. We recall that each filter was analyzed independently.

\subsection{Normalized amplitudes across wavelengths}\label{subsec:Normalized Amplitudes Across Wavelengths}  % 4.2
% Table 3
\begin{table}%[]
    \caption{Normalized amplitudes with their interquartile range (IQR; $p_{75} - p_{25}$).}
    \label{tab:Percentile Errors in the Medians of the Normalized Amplitudes}
    \centering
    \begin{tabular}{p{1.5cm}p{2.5cm}}
    \hline\hline
      Passband & $\overline{A}$ \\
      \hline 
      $J0395$ & $1.73 \pm 0.32$ \\
      $g$ & $1.36 \pm 0.14$ \\
      $J0515$ & $1.25 \pm 0.13$ \\
      $r$ & $1.00$ \\
      $J0660$ & $0.88 \pm 0.09$ \\
      $i$ & $0.79 \pm 0.12$ \\
      $J0861$ & $0.73 \pm 0.17$ \\
      \hline
    \end{tabular}
\end{table}
%% Text
To quantitatively study the observed trend of decreasing amplitude with increasing wavelength, the normalized amplitude with respect to the $r$- band amplitude was computed, noted $\overline{A}$ (see Fig.~\ref{fig:normApli_vs_wavelengths}). This permits removing the main trends present in the Bailey Diagrams and enhances relative differences across passbands. The interquartile ranges (IQR; defined as the difference between the 75\textsuperscript{th} and 25\textsuperscript{th} percentiles, $p_{75} - p_{25}$) of the normalized amplitudes, along with their medians, are summarized in Tab.~\ref{tab:Percentile Errors in the Medians of the Normalized Amplitudes} for all seven optical bands in J-VAR DR1. A clear downward trend was observed for increasing wavelengths, indicating that normalized amplitude decreases as wavelength increases, suggesting that the fluxes in bluer bands are more sensitive to temperature variations than those in the redder bands.

An exponential function was fitted to the medians of normalized amplitude ratios across the J-VAR DR1 bands. The resulting best-fit curve was given by
\begin{equation}
\overline{A}(\lambda) = 7.78 \cdot e^{-0.005 \cdot \lambda} + 0.61,
\label{eq:amplitude_lambda}
\end{equation}
where $\overline{A}(\lambda)$ is the normalized amplitude and $\lambda$ is the wavelength in nanometers. This relation captures the expected physical behavior of RR Lyrae stars, where variability amplitude diminishes at longer wavelengths due to the lower temperature sensitivity of the flux described by the Planck Function \citep{planck1901}.

\subsection{Residuals}
\label{subsec:Residuals}% 4.3
\begin{table}%[] % Table 4
    \caption{Gaussian fit parameters of residuals for each band.}
    \label{tab:gaussian_fit_parameters}
    \centering
    \begin{tabular}{p{1.5cm}cc}
    \hline\hline
      Band & Mean & Standard Deviation \\
      \hline
      $J0395$ & 0.12 & 1.19 \\
      $g$ & 0.06 & 1.37 \\
      $J0515$ & 0.06 & 1.14 \\
      $r$ & 0.03 & 1.25 \\
      $J0660$ & 0.05 & 1.12 \\
      $i$ & 0.05 & 1.17 \\
      $J0861$ & 0.03 & 1.09 \\
    \hline
    \end{tabular}
\end{table}

%%% Text
Residuals were defined as the differences between data and fits, and the ratio of residuals to magnitude errors represents the units of error by which the fit differs from the data. This analysis is crucial for assessing the quality of the fit. Figure \ref{fig:hist_residuals} shows the distribution of residuals normalized by magnitude errors across the seven optical bands in J-VAR DR1, with a Gaussian fit overlaid for comparison. The mean and standard deviation of the Gaussian were estimated separately for each filter and are presented in Tab.~\ref{tab:gaussian_fit_parameters}.

The near-zero mean in the residual distributions indicates no significant systematic bias in the template fits. The expected standard deviation in an ideal case is unity, and larger values around $1.2$ were found. Two origins are possible: the J-VAR photometric errors are underestimated or the SDSS templates do not reflect all the features present in the observational light curves. A combination of both effects is also possible. Interestingly, the unique J-VAR passbands $J0395$, $J0515$, $J0660$, and $J0861$ do not present larger standard deviations than the broadbands, suggesting that the SDSS template library provides a proper description of the RR Lyrae light curves in these filters within the limitations already discussed. 

The estimation of empirical light curves from J-VAR data or the use of different libraries (e.g. \citealt{baeza25}) is needed to get additional clues about the origin of the enhanced dispersion, highlighting important areas for improvement in template libraries and error modeling in variable star photometry. This is beyond the scope of the present paper and will be addressed in future work.

%%%%%%%%%
% Section 5
\section{Discussion}\label{sec:Discussion}
% figure xx Agreement Rate
\begin{figure*}
    \centering
    \begin{subfigure}[]{0.7\textwidth}
        \centering
        \includegraphics[width=\textwidth]{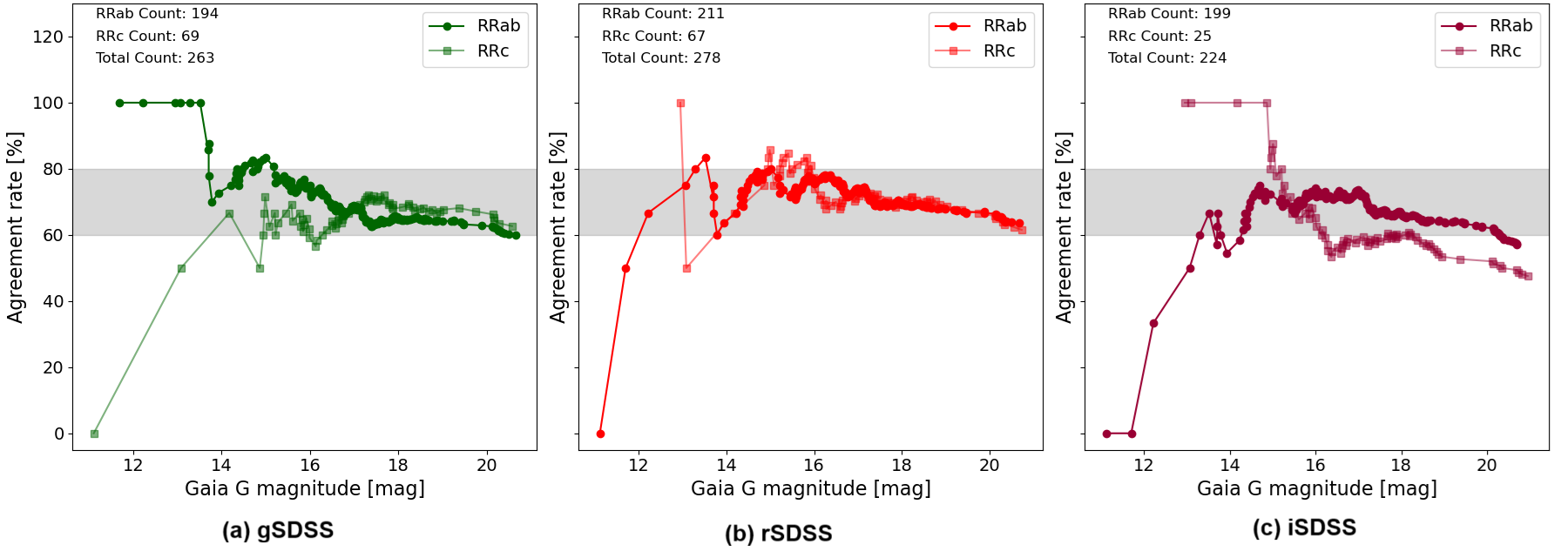}
        \label{fig:gri_success_rate}
    \end{subfigure}
    %\hfill
    \begin{subfigure}[]{1.\textwidth}
        \centering
        \includegraphics[width=\textwidth]{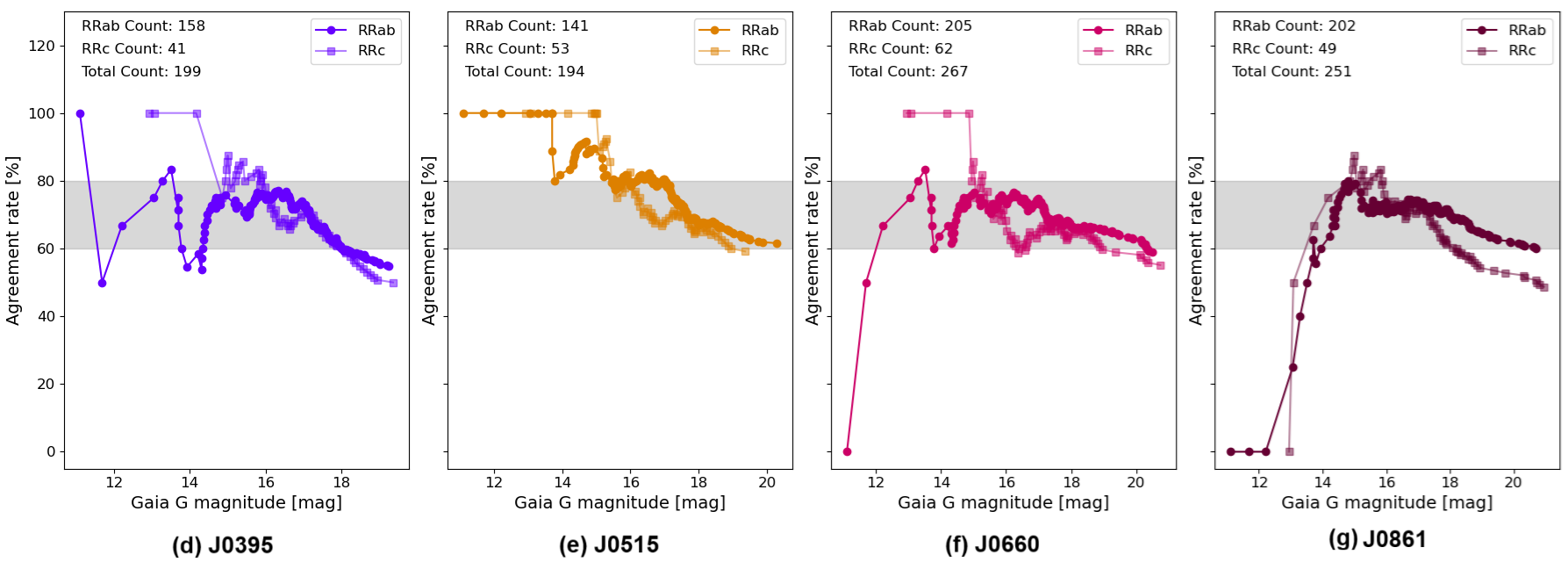}
        \label{fig:jvarbands_success_rate}
    \end{subfigure}
    \caption{Agreement rates of period estimates between J-VAR and \textit{Gaia} as a function of \textit{Gaia} $G$-band magnitude: This figure shows the percentage of RR Lyrae stars whose J-VAR periods agree with \textit{Gaia} periods within 0.01 days, plotted against their \textit{Gaia} $G$-band magnitude. The shaded gray region highlights agreement rates between 60\% and 80\%. RRab and RRc counts indicate the number of stars within this range for each subtype, along with the total number of stars analyzed.}
    \label{fig:success_rates}
\end{figure*}

%% TEXT %%%%%%%
\subsection{Comparison of periods} %5.1
\label{subsec:Comparison of periods}
The reliability of the J-VAR periods was assessed by comparing them with the periods in the \textit{Gaia} DR3 Variable Star catalog. For each source, the absolute differences between the J-VAR DR1 periods ($P_{J-VAR}$) in days and the \textit{Gaia} DR3 periods ($P_{Gaia}$) were computed as
\begin{equation}
\Delta P = \left| P_{\mathrm{J-VAR}} - P_{\textit{Gaia}} \right|.
\label{eq:deltaP}
\end{equation}
A threshold of $\Delta P < 0.01$ days was adopted to evaluate the agreement between the two datasets (Fig.~\ref{fig:success_rates}). The figure presents the fraction of J-VAR periods that agree with the Gaia values within our adopted threshold, shown as a function of the Gaia $G$ magnitude. Across all bands, period agreement rates peak near $15$ mag, with RRab stars achieving roughly $80$\% consistency and RRc stars performing even better at around $90$\%. The agreement rate drops for fainter stars, with RRab stars gradually decreasing, while RRc stars decline more quickly. For brighter stars, the agreement rate drops sharply. This is partly due to saturation effects in stars observed during exceptionally clear, photometric nights, where very high signal-to-noise ratios can cause non-linear detector response. Zero-agreement cases may result from sparse data or limitations in the SDSS multiband template library, which may not fit well across all seven optical bands in J-VAR. For a threshold of $0.01$ days, RRab stars exhibit higher agreement rates than RRc stars, suggesting that fundamental mode pulsators were more reliably modeled by the template fitting method.

The SDSS multiband template library provides a more comprehensive understanding of RR Lyrae stars by integrating data from various bands. This allows for improved classification, more accurate period determination, and better insights into pulsation processes, e.g., mass loss, binary interactions \citep{Bobrick_2024}. Moreover, a recent study \citep{Ngeow_2025} includes a template library for $y$-band in the existing SDSS $ugriz$ bands. This inclusion provides an extended wavelength range, especially at longer wavelengths, improves the overall fit of the light curves, and facilitates a better classification of RR Lyrae subclasses. This is considered essential for understanding stellar evolution within our galaxy, particularly in populations such as the Galactic halo and globular clusters. By determining the subtype RRab, RRc, or RRd, more accurate estimations of the age and metallicity of these systems can be made. The creation of additional template libraries for the J-VAR medium and narrow bands is beyond the scope of the present work, but the residual analysis in Sect.~\ref{subsec:Residuals} suggests that the current broadband templates capture most of the features present in the data.

\subsection{Amplitude trends in the \textit{Gaia} and ZTF surveys} %5.2
\label{subsec:Amplitude Trends}
The normalized amplitude in the \textit{Gaia} $G$-band, and the ZTF $g$ and $r$ bands were computed to compare with the observed J-VAR trend. To minimize selection effects, only the RR Lyrae stars in the J-VAR DR1 catalog were used. A pivot wavelength of $622.88~\text{nm}$ for the \textit{Gaia} $G$-band was assumed \citep{Weller2018}, while the ZTF $g$ and $r$ bands share profiles similar to those of J-VAR.

The measured normalized amplitudes, shown in Fig.~\ref{fig:normApli_vs_wavelengths}, were $\overline{A}\,(G) = 1.08 \pm 0.21$ for {\it Gaia}, and $\overline{A}\,(g) = 1.38 \pm 0.22$ and $\overline{A}\,(r) = 0.98 \pm 0.15$ for ZTF. The normalized amplitude of the \textit{Gaia} $G$-band follows the J-VAR exponential trend; it lies slightly above the exponential curve. This deviation stems from the broad nature of the \textit{Gaia} $G$-band, which spans both blue and red wavelengths. Since most RR Lyrae stars are intrinsically blue, the $G$-band, despite its redder effective wavelength, still captures significant flux from the blue spectrum. Moreover, \textit{Gaia} $G$-band magnitudes derive from fluxes convolved over a wide spectral range, introducing a non-linear response compared to narrower bands. This broad convolution may enhance measured variability relative to what a narrow, red band would detect, causing the slightly elevated median amplitude \citep{Jordi2010}.

In contrast, the ZTF $g$ and $r$ bands align with the corresponding J-VAR passbands, allowing direct comparison across surveys. A near-identical placement of ZTF and J-VAR $g$ and $r$ band points along the exponential amplitude–wavelength trend is found, reinforcing our results.

%%%%%
\section{Conclusions}
\label{sec:Conclusions}
This study presents the J-VAR DR1 + {\it Gaia} DR3 catalog, comprising $315$ RR Lyrae stars selected by cross-matching J-VAR DR1 with the {\it Gaia} DR3 Variable Star catalog. The J-VAR DR1 + {\it Gaia} DR3 catalog includes J-VAR photometry in seven optical bands and derived parameters from the light curve analysis.

The RR Lyrae light curves in each J-VAR passband were analyzed by fitting templates from the SDSS multiband template library to obtain periods and pulsation amplitudes. J-VAR periods showed strong agreement with those in the \textit{Gaia} DR3, validating the accuracy of the template-fitting method. Discrepancies observed in a small subset of objects were likely due to low signal-to-noise or sparse sampling. The best-fitting residuals in the $J0395$, $J0515$, $J0660$, and $J0861$ passbands are similar to those in the $gri$ broadbands, reflecting that SDSS broadband templates can also reproduce the observations in the medium and narrow bands.

The Bailey diagrams constructed using J-VAR DR1 periods and amplitudes successfully recover the expected sequences for RRab and RRc stars in all seven optical bands in J-VAR. The known trend with wavelength is recovered, with bluer passbands having larger amplitudes than redder passbands \citep[e.g.][]{Catelan2009}. This trend was quantified with an exponential function after normalization to the $r$-band amplitudes, providing a benchmark for theoretical models. The normalized amplitudes provided by {\it Gaia} and ZTF for common RR Lyrae stars are compatible with the J-VAR results.

The resulting J-VAR DR1 + \textit{Gaia} DR3 catalog stands as a valuable asset, poised to support ongoing and future research into RR Lyrae stars. The inclusion of narrow bands, particularly H$\alpha$ ($J0660$) and \ion{Ca}{ii} H \& K-sensitive $J0395$, opens new possibilities to investigate chromospheric activity, magnetic fields, and pulsation-driven phenomena. Though no significant phase shifts were found in the current J-VAR DR1 sample, future $J0660$ and $J0395$ analyses may uncover subtle atmospheric dynamics.

This study also sets the stage for exploring links between RR Lyrae and Blue Horizontal Branch stars, shedding light on shared evolutionary paths and the Milky Way’s formation history \citep{Carretta2004}. The methods developed here may be extended to RR Lyrae populations in other Local Group galaxies, enhancing models of stellar evolution.
%%%%%%%

\section*{Acknowledgements}

S.~K. has received financial support from the Aragonese Government through the Research Groups E16\_23R and the Spanish Ministry of Science and Innovation (MCIN/AEI/10.13039/501100011033 y FEDER, Una manera de hacer Europa) with grant PID2021-124918NB-C41.

A.~E. acknowledges the financial support from the Spanish Ministry of Science and Innovation and the European Union - NextGenerationEU through the Recovery and Resilience Facility project ICTS-MRR-2021-03-CEFCA.

F.~J-E. acknowledges that this research has been partially funded by MICIU/AEI/10.13039/501100011033/ through grant PID2023-146210NB-I00.

Based on observations made with the JAST80 telescope and T80Cam camera for the J-PLUS project at the Observatorio Astrof\'{\i}sico de Javalambre (OAJ), in Teruel, owned, managed, and operated by the Centro de Estudios de F\'{\i}sica del  Cosmos de Arag\'on (CEFCA). We acknowledge the OAJ Data Processing and Archiving Department (UPAD; \citealt{upad}) for reducing and calibrating the OAJ data used in this work, as well as for the publication of the data products through a dedicated web portal.

Funding for OAJ, UPAD, and CEFCA has been provided by the Governments of Spain and Arag\'on through the Fondo de Inversiones de Teruel and their general budgets; the Aragonese Government through the Research Groups E96, E103, E16\_17R, E16\_20R and E16\_23R; the Spanish Ministry of Science and Innovation (MCIN/AEI/10.13039/501100011033 y FEDER, Una manera de hacer Europa) with grants PID2021-124918NB-C41, PID2021-124918NB-C42, PID2021-124918NA-C43, and PID2021-124918NB-C44; the Spanish Ministry of Science, Innovation and Universities (MCIU/AEI/FEDER, UE) with grant PGC2018-097585-B-C21; the Spanish Ministry of Economy and Competitiveness (MINECO) under AYA2015-66211-C2-1-P, AYA2015-66211-C2-2, AYA2012-30789, and ICTS-2009-14; and European FEDER funding (FCDD10-4E-867, FCDD13-4E-2685).

We thank the J-VAR Collaboration, as well as Álvaro Álvarez-Candal and Vinicius Placco of the J-PLUS Collaboration, for their valuable comments.

%%%%%%%%%%%%%%%%%%%%%%%%%%%%%%%%%%%%%%%%%%%%%%%%%%
\section*{Data Availability}
The J-VAR DR1 data used are publicly available at CEFCA's catalog portal\footnote{\url{https://archive.cefca.es/catalogues/jvar-dr1}}. The derived parameters for the $315$ RR Lyrae studied in this work are available at Zenodo\footnote{\url{https://zenodo.org/records/17601696}}.


\bibliographystyle{aa}
\bibliography{references} % if your bibtex file is called example.bib
%\begin{thebibliography}{99}
%\end{thebibliography}

%%%%%%%%%%%%%%%%% APPENDICES %%%%%%%%%%%%%%%%%%%%%

%\appendix

\begin{appendix}
    
\section{J-VAR DR1 RR Lyrae Catalog}
\label{App:J-VAR DR1 RR Lyrae Catalog}
This appendix describes the J-VAR DR1 + \textit{Gaia} DR3 catalog of light curve analyses for $315$ RR Lyrae stars. The catalog includes both observational data and derived pulsation parameters. Tab.~\ref{tab:Nr._of_objects} summarizes the number of sources with available periods and amplitudes in each of the seven J-VAR optical bands. Tab.~\ref{tab:rrlyrae_catalog} provides a detailed overview of the catalog contents.

\begin{table}%[h]
    \caption{Number of RR Lyrae stars fitted for each configuration of J-VAR DR1 bands mapped to the SDSS multiband template library.}
    \label{tab:Nr._of_objects}
    \centering
    \begin{tabular}{p{2cm}p{2cm}p{2cm}}
    \hline\hline
      J-VAR DR1 Band & SDSS Template Band & No. of Stars Fitted\\
      \hline
      $J0395$ & \textit{ug}  & 272 \\
       \textit{g} & \textit{g}  & 303 \\
       $J0515$ & \textit{g}  & 283 \\
       \textit{r}  & \textit{r}  & 308 \\
       $J0660$ & \textit{r}  & 304 \\
       \textit{i}  & \textit{i}  & 313 \\
       $J0861$ & \textit{z}  & 298 \\
       Total & - & 2081\\
       \hline
    \end{tabular}
\end{table}
%\FloatBarrier

\begin{table}%[h]
    \caption{Description of the J-VAR DR1 $+$ \textit{Gaia} DR3 catalog of light curve analysis of $315$ RR Lyrae.}
    \label{tab:rrlyrae_catalog}
    \centering
    \begin{tabular}{p{2.5cm}p{5.5cm}}
        \hline\hline
        Attribute & Description \\
        \hline
        obj\_id & Unique J-VAR DR1 Identifier \\
        gaia\_sid & \textit{Gaia} Source Identifier, linking it the star to \textit{Gaia} DR3 Variable Star catalog \\
        star\_class & Classification of the RR Lyrae star (e.g., RRab, RRc)\\
        RA, Dec & Right Ascension and Declination (J2010 epoch) \\
        gaia\_pf & \textit{Gaia} DR3 periods (in days) in fundamental mode of frequency, retrieved from the \textit{Gaia} DR3 Variable Star catalog \\
        gaia\_p1o & \textit{Gaia} DR3 periods (in days) in the first overtone, retrieved from the \textit{Gaia} DR3 Variable Star catalog   \\
        templ\_fit & Best-fitted template in the SDSS multiband template library\\
        jvar\_period & Periods (in days) of the best-fitted templates \\
        gaia\_amplitudes & \textit{Gaia} amplitudes (in mag) computed as halves of peak-to-peak \textit{Gaia} $G$-band magnitudes\\
        jvar\_amplitudes & J-VAR amplitudes (in mag) computed as half of the difference between the minimum and maximum of the J-VAR magnitudes. \\
        \hline
    \end{tabular}
\end{table}
%\FloatBarrier

\end{appendix}

%%%%%%%%%%%%%%%%%%%%%%%%%%%%%%%%%%%%%%%%%%%%%%%%%%

% Don't change these lines
%\bsp	% typesetting comment
\label{lastpage}
\end{document}